# Simulation and Analysis of Quality of Service (QoS) Parameters of Voice over IP (VoIP) Traffic through Heterogeneous Networks

Mahdi H. Miraz
School of Computer Studies
AMA International University BAHRAIN
Salmabad,Bahrain
Centre for Ultra-realistic Imaging (CURI)
Glyndŵr University, UK

Muzafar A. Ganie
Department of Computer Science &
Software Engineering University of Hail,
Hail, KSA
KSA

Suhail A. Molvi
Department of Computer Science &
Software Engineering, University of Hail,
Hail, KSA

Maaruf Ali
Department of Science and Technology,
University of Suffolk,
Ipswich, Suffolk, UK

AbdelRahman H. Hussein
Department of Software Engineering,
Al-Ahliyya Amman University,
Amman, Jordan

*Abstract*—**Identifying those causes and parameters that affect the Quality of Service (QoS) of Voice-over-Internet Protocol (VoIP) through heterogeneous networks such as WiFi, WiMAX and between them are carried out using the OPNET simulation tool. Optimization of the network for both intra- and inter-system traffic to mitigate the deterioration of the QoS are discussed. The average value of the jitter of the VoIP traffic traversing through the WiFi-WiMAX network was observed to be higher than that of utilizing WiFi alone at some points in time. It is routinely surmised to be less than that of transiting across the WiFi network only and obviously higher than passing through the increased bandwidth network of WiMAX. Moreover, both the values of the packet end-to-end delay and the Mean Opinion Score (MOS) were considerably higher than expected. The consequences of this optimization, leading to a solution, which can ameliorate the QoS over these networks are analyzed and offered as the conclusion of this ongoing research.**

## I. INTRODUCTION

Because of the ever increasing and global adoration of using the Internet, especially for Voice-over-IP (VoIP) calls on mobile devices, it is turning out to be progressively inexpedient to disregard the gravity of voice communications utilizing the Internet in our everyday lives. Due to the continuance of dissimilar types of protocols and networks (i.e. WiFi, WiMAX, 3G, 4G, LTE, CDMA, GSM, EDGE, GPRS etc.), in most cases the data has to traverse multiple assorted networks - there is an urgent need for this research. While VoIP traffic passes through any such heterogeneous networks, the Quality of Services (QoS) suffers noticeable degradation. The solitary raison d'être of the research, presented in this paper, is to explore and investigate the level and magnitude of such degradation of the QoS of VoIP traffic traveling through these assorted networks. In pursuance of this aim, our objectives are of threefold: 1) to design, develop and configure appropriate sample networks using the OPNET modeler; 2) to run the simulation using various loads as well as to record the measured results of the QoS parameters; and finally 3) to articulate the research findings by analyzing the results procured through the simulations. The first two scenarios are made up of a number of VOIP clients transferring data through a couple of homogeneous networks i.e. WiMAX-to-WiMAX, WiFi-to-WiFi. The major QoS parameters of VoIP traffic such as the: Mean Opinion Score (MOS), Throughput, Availability, Crosstalk, Jitter, Distortion, Link Utilization Distribution, Attenuation, Loss and Echo, etc. are to be scanned and analyzed. The third set-up comprises of heterogeneous networks replacing the homogeneous ones. The VoIP traffic traverse a heterogeneous network made up of assorted protocols i.e. WiMAX-to-WiFi. The simulation will capture the same VoIP QoS parameters as in the first couple of scenarios. The results, thus obtained using the heterogeneous networks, will then be analyzed and compared with the previously attained results using the homogeneous set-ups.

The layout of this research comprises the arrangement of the following parts: The first section imparts a concise preamble to the research while the second section gives a detailed account of the background information as well as the relevant technological/scientific terms referred to in this report. The third section comprises a "Literature Review" survey studying a broad selection of research projects and articles whereas the fourth section covers the research methodology together with the simulation scenarios of the networks as well as the necessary configuration/set-up to accomplish them. The fifth section analyzes and compares the results, followed by the concluding discussion together with the layout for potential future research directions and works.

## II. BACKGROUND TERMINOLOGY

### A. Voice over IP (VoIP)

Voice over Internet Protocol or more commonly known as "VoIP" [1] is simply defined as the digitized voice traffic intrinsically transmitted using a data network to make telephone calls. This differs from using a traditional analogue circuit switched public network, as now the data has been split into packets. These packets can take any route to reach the destination. Packetized data travel through a virtual circuit which differs from a circuit switched network in that the circuit does not need to be reserved for the entire duration of the call between the sender and the receiver with packet switching. Thus the channels may be utilized more by sharing with other users than compared to circuit switching. However, the data packet can arrive out of sequence, experience delay or even may never arrive as a consequence of traffic congestion and buffer overflows. These are some of the major disadvantages of sharing traffic across a virtual network that VoIP traffic has to contend with. On the other hand, the advantages offered include the multiple routing of the VoIP traffic ensuring a cheaper and often free of cost flow of traffic between the different intra-packet network components such as the routers and switches. Transmitting digital data in the format of packets signifies that all types of digitized data such as voice, video, fax, music and telephony have the opportunity to be carried together utilizing a shared common network at any given time.

The fact of being software packet based puts VoIP technology in a favourable or superior position. Thus, VoIP enjoys a distinct advantage and supremacy of budget scalability in comparison with the currently operational alternative telephony systems. This allows lines to be shared with other users and services thus helping to lower the overall costs over the circuit switched networks. However, being predominantly a network based on software - it is exposed to the possibility of being attacked or harmed by the progressively rising threat of cyber-attacks from crackers in terms of malware such as viruses and worms. In [2], the author discusses several security solutions to confront this potential problem.

Convergence has been accelerated with the deployment of 3G [3], WiMAX and considerably further recently by the deployment of LTE and 4G, particularly amongst internet, mobile and fixed services. Universal access to the internet regardless of the means of transportation is accelerating predominantly due to the widespread rollout of WiMAX, WiFi and femtocells in public spaces. The demand for greater bandwidth to support multimedia broadband access is also increasing and being expected by the consumers. This was facilitated by the adoption of the IP Multimedia Subsystem (IMS) in the Rel. 5 version of UMTS (Universal Mobile Telecommunications System). The IMS is a packet based control overlay network used for transporting user data and signaling.

The Session Initiation Protocol (SIP), a development of the Internet Engineering Task Force (IETF) was embraced by the Third Generation Partnership Project (3GPP) for setting up IP-based multimedia sessions, this includes VoIP. The current IEEE 802.11 (WiFi) and 802.16 (WiMAX) networks completely support VoIP and many other real-time services [4].

### B. Session Initiation Protocol (SIP)

Making, maintaining and clearing a call requires control information and signaling to be exchanged between the network entities. This is actually a rather complicated process where internet mobility is involved across various types of devices with differing capabilities and network technologies. A protocol that has been chosen to manage these "sessions" is known appropriately as the "Session Initiation Protocol" or SIP [5]. SIP works alongside and in complement with the existing real-time protocols. The source and destination endpoints, known as the "user agents", discover each other and then negotiate the parameters for the efficient exchange of information by the use of SIP. The necessary user agents and intermediary nodes are handled by SIP by the creation of proxy servers. These proxy servers can then request and respond to 'invitation', 'registration' and other such SIP requests. SIP is a transport protocol independent of the type of session being established. SIP is designed to be agile, flexible and to handle various types of multimedia data exchange.

SIP being an application layer control protocol can take care of the entire multimedia call set-up to the termination process. It also includes the ability to handle multicast call set-up, including the removal of the participant. SIP is designed for mobility with features such as redirection and name mapping. A powerful feature of SIP is the ability to maintain an externally visible identifier, invariant of location [6]. For example, SIP supports these call set-up features: session set-up, session management, user availability, user location and user capabilities.

### C. QoS Parameters of VoIP Traffic

The data networks being flexible in its ability to handle multifarious types of data services over the Public Switched Telephone Network (PSTN) puts the Plain Old Telephone Service (POTS) at a financial disadvantage [4]. The QoS parameter of VoIP traffic varies, and can be quantified by a range of divergent metrics, such as the: jitter, end-to-end delay and Mean Opinion Score (MOS), as shown in Table 1.

The Mean Opinion Score (MOS) has been used to subjectively measure the voice quality in a telephone network. It is based on a perceptual scale of 1 to 5 as shown in Table 1.

TABLE. I.  SCALING AND CLASSIFICATION OF MOS [7].

| Score | Quality | Scale of Listening Effort |
|---|---|---|
| 5 | Excellent | No effort is required. |
| 4 | Good | No considerable effort is required. |
| 3 | Fair | Moderate effort is required. |
| 2 | Poor | Considerable effort is required. |
| 1 | Bad | Not understood even with considerable effort. |

Jitter "is the variation in [the] arrival time of consecutive packets" [10]. Jitter is calculated over an interval of time [7]. It should be noted that the buffers can both under-fill and over-fill, triggering packet drops.

The packet end-to-end (E2E) delay "is measured by calculating the delay from the speaker to the receiver [including the] compression and decompression delays" [8].

The International Telecommunication Union – Telecommunication (ITU-T) gives the guidelines for the delay and jitter for the different types of call quality, as presented in Table 2 [8].

TABLE. II.  ITU-T PRECEPT FOR VOICE QUALITY [8].

| Network Parameter | Good | Acceptable | Poor |
|---|---|---|---|
| **Delay (ms)** | 0-150 | 150–300 | > 300 |
| **Jitter (ms)** | 0-20 | 20–50 | > 50 |

### D. WiFi™ (IEEE 802.11x)

The contention wireless networking technology, WiFi, evolved from its counterpart wired IEEE Ethernet 802.3, outlining perceptions for the technology of Local Area Network (LAN), to become the IEEE 802.11 Wireless LAN or WLAN. The physical and data link layers are defined, operating over the two different frequency bands of 2.4 GHz and 5 GHz. Two popular WiFi standards are the 802.11b (11 Mbit/s) and the 802.11g (54 Mbit/s) with an operating range of 80-100 m. The protocols being a contention based system, the speeds quoted are a theoretical maximum. The contention causes the comparatively low bitrates and thus affects the QoS, especially for real-time services like VoIP. This is not helped by the large headers of the WiFi and VoIP protocols themselves. Its uptake and popularity has been due to the inexpensive price of the router and most network equipment coming with its built-in, including the WiFi antenna. WiFi has now become widespread covering: domestic, industrial, public spaces including on public transportation [9].

### E. WiMAX™ (Worldwide Interoperability for Microwave Access) Technology

WiMAX, when it was first introduced ten years ago was meant to provide a global wireless high speed mobile Internet access. However, LTE (Long Term Evolution) has largely superseded this application. WiMAX, however, is not dead and there are around 580 operators in the world providing backhaul and rural access to fast broadband internet access, often in the less developed regions of the world. Typical application scenarios of WiMAX are shown in Fig. 1. WiMAX was designed to provide the same experience as that of fixed internet services, such as QoS, Service Level Agreement (SLA), interoperability with off course mobility, wide coverage and security [10]. It is ironic that WiMAX, once touted as the "4G of Wireless Technology" has now been superseded ahead of its time by LTE. WiMAX is still probably the first all IP mobile internet technology allowing true scalability to carry multimedia traffic [11]. WiMAX provides a coverage area of 50 km$^2$ with data rates of 75 Mbps [12].

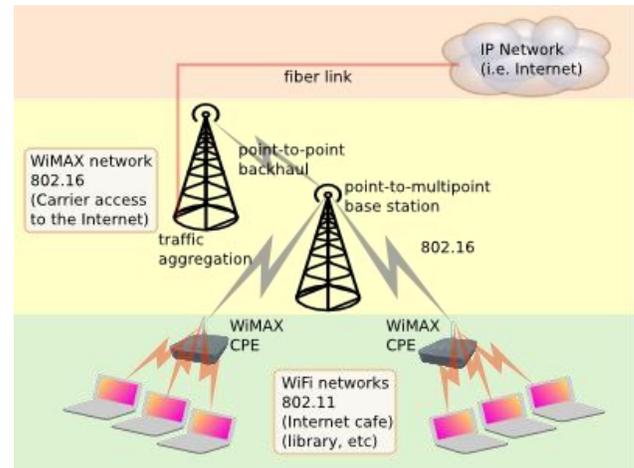

Fig. 1.  Application Scenarios of WiMAX. (From: http://www.accessmillennium.com/images/wifi_vs_wimax.png)

WiMAX comes in two types of technologies: the fixed IEEE 802.16/a/d version and the wireless IEEE 802.16-2005 (16e) amendment [13]. The latest version is known as WiMAX rel 2 or IEEE 802.16m. The latest version allows download bitrates up to 1 Gbit/s through channel aggregation for low mobility users.

## III. LITERATURE REVIEW

In another study Mahdi *et al.* [14], [15] investigated the same QoS parameters but for VoIP traffic travelling through UMTS and WiFi alone and together.

A previous simulation study of VoIP over both WiFi and WiMAX [9] has shown that VoIP activity does impact negatively on the overall throughput of both technologies. However, only in the WiFi network is packet loss and jitter experienced. The parameters commonly used to study the performance of the network, for example a study of WiMAX and UMTS using the OPNET network simulation software include: "MOS, end-to-end delay, jitter, and packet delay variation" [7].

It would appear that not all software implementations of VoIP clients are equal - as they vary in their effect on voice quality. This was revealed by a research experiment performed over the High Speed Packet Access (HSPA) [14] service.

To succeed in dealing with the severe problems of VoIP calls over WiFi while approaching the WiFi capacity limit and congestion, a new scheme, the Quality Assurance of Voice over WLANs (SQoSMA) [16] was proposed. SQoSMA took the approach of incorporating the data with the control and planes for detecting and mitigating congestion events. This was achieved by selecting the appropriate adaptive audio codec with the suitable bitrate and then implementing a call stopping method where needed to fix congestions.

An earlier similar scheme [17] was also explained with the use of edge VoIP gateway between the WLAN and the Internet Cloud. The task of the edge VoIP gateway was to determine the pertinent variable speech coding rate (64, 40, 32, 24 and 16 Kbit/s) to lessen the network congestion with a subsequent increase in the overall QoS of speech traffic.

A technique that reduces VoIP traffic's packetization delay (also known as transmission delay or store-and-forward delay) utilized a Transmission Control Protocol (TCP), Friendly Rate Control (TFRC) algorithm based 802.11e network which applied the EDCF (Enhanced Distributed Coordination Function)/HCF (Hybrid Coordination Function) scheme [17].

In [18], authors proposed using a routing and label based solution for transporting real-time VoIP traffic through WLAN which efficiently processed the procedures of call QoS, mobility and call admission. Their procedure utilized a 15 node wireless mesh network to implement distributive packet aggregation utilizing MAC waiting without unbounded packet delays. The fully optimized procedure resulted in a performance gain of 13 times for six hops.

Since human voice is assessed by humans and is therefore purely subjective, a metric to assess this for VoIP traffic is needed that takes into account human subjectivity — which is lacking in the purely objective SNR (signal-to-noise ratio) measure. A study [19] in this field was conducted to look at such metrics concentrating on the E-Model and the Perceptual Evaluation of Speech Quality (PESQ). The researchers studied the limitations of both measures and devised a new metric consolidating the advantages and benefits of them to devise the Advanced Model for Perceptual Evaluation of Speech Quality (AdmPESQ). AdmPESQ is particularly applicable for heterogeneous types of networks with differing delay parameters and packet losses.

The popularity of VoIP has been mushrooming since the last few years. VoIP is now routinely utilized by a wide range of diverse populations globally. While lowering the call price rates, VoIP facilitates almost all the advantages offered by the traditional Public Switched Telephone Network (PSTN). Furthermore, it incorporates several additional value added features. As a consequence of its widespread popularity and such advantages, many companies penetrated into the business of offering various VoIP services. The VoIP traffic, thus, has to pass across several different types of networks — often heterogeneous in nature. Degradation of Quality of Service (QoS) was thus experienced whilst the traffic traverses across such assorted networks. Materna [20], in his research paper "VoIP insecurity", has enumerated four types of attacks that are relevant to VoIP, viz.:

- Eavesdropping;
- Service integrity;
- Service availability; and
- Spam over Internet Telephony (SPIT).

The successful availability without network outage is vital for the success on any well networked and connected corporation. Thus protection against any forms of "Service Availability Attacks" is of paramount importance. Downtime in the telephony network will mean: lost revenues for the enterprise and the service providers, unplanned maintenance costs and lost productivity. The IP Telephony network must be protected against all known forms of attacks, which include: viruses, worms and especially the variations of "Denial of Service" (DOS). The effects of these may range from the degradation of the QoS to the total loss (also known as call drops) of the service. Degradation of the QoS is not just a minor nuisance but actually of major concern as customers often request the highest voice quality when they subscribe to an IP Telephony service.

The effect of such an attack on VoIP is actually more sensitive and harmful as it has a lower threshold and immunity than computer data networks. Computer data networks are protected more securely and are usually affected to a lesser degree than the VoIP network. Thus a generic worm may adversely affect the VoIP network precisely because of these reasons, in advance of the computer network. The worm may at most, just slow down the computer data network. The worm may, however, totally bring the VoIP network down.

The aim of this research is to ascertain the degree to which the VoIP traffic's quality of service (QoS) deteriorates while traversing through heterogeneous networks. In order to achieve this aim, the authors of this paper, carefully designed, developed and simulated several network scenarios using the OPNET modeler. The results of the various VoIP QoS parameters, thus obtained through the simulation, were then analyzed, reported and published in the literature.

## IV. RESEARCH METHOD

Due to financial constraints and equipment limitations, the simulation of a sample network, especially in academic research, is very important in the fields of computer networking and telecommunication. Not only does it help to get the perspective view of a network, it also provides guidance for the future. Jack Burbank [21] describes "Modeling and Simulation (M&S)" as an acute constituent in the "design, development and test and evaluation (T&E)" process. As reported by him, "It is almost always preferable to have [an] insight into how a particular system, individual device, or algorithm will [actually] behave and perform in the real world prior to its actual development and deployment" [21]. The advantages of M&S take account of the capability of exercising scenarios and case-studies which are not easily achievable through any empirical methods such as: network scalability testing; the capacity to adapt models to test the systems' sensitivity and to tune its performance [22]. In the case of two or more similar available technologies, it helps to compare and contrast in order to take deployment decisions. This project utilizes and takes advantages of the OPNET Modeler simulation software because it effectively incorporates a wide variety of protocols and technologies [23] while comprising a "development environment". This smoothes the process of M&S of different types of networks and technologies including (but not limited to): VoIP, WiMAX, WiFi, 3G and LTE. Other networking technologies can be written in software or are available from third party sources.

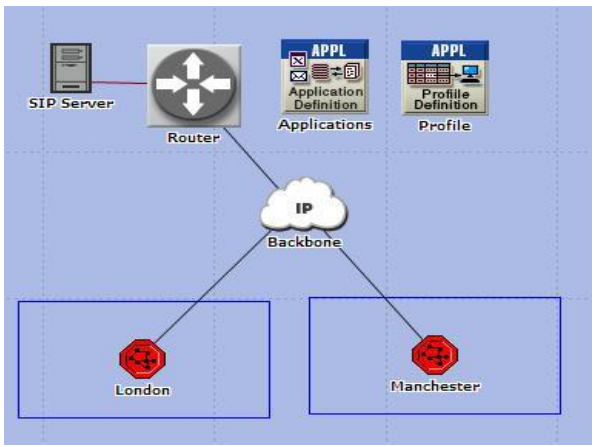

Fig. 2. WiFi network scenario.

In our first simulation scenario, a pair of WiFi subnets, namely London and Manchester, was designed and deployed. As shown in Fig. 2, both the subnets are configured with SIP server credentials connected via an IP cloud.

In our second simulation scenario, a pair of WiMAX subnets, namely Cambridge and Bradford was deployed instead of the WiFi ones. The last scenario replaces one of the WiMAX subnets (namely Bradford) from the second scenario by one of the WiFi subnets (namely Manchester) from the first scenario. Table 3 illustrates some details of the subnets deployed in this research project:

TABLE. III.    LIST OF DEVICES USED CONFIGURING THE SUBNETS

| Subnet Name | Scenario | Base Station Type | Work Station Type | Number of Work Stations |
|---|---|---|---|---|
| **London** | WiFi | WiFi | Mobile | 4 |
| **Manchester** | WiFi | WiFi | Mobile | 4 |
| **Cambridge** | WiMAX | WiMAX | WiMAX Workstation | 4 |
| **Bradford** | WiMAX | WiMAX | WiMAX Workstation | 4 |
| **Manchester** | WiMAX_Wi Fi | WiFi | Mobile | 4 |
| **Cambridge** | WiMAX_Wi Fi | WiMAX | WiMAX Workstation | 4 |

It was obviously possible to add more workstations to the scenario, however, we were not interested in the network load, network complexity or routing. Rather, the aim of the research is to find the degradation of the QoS due to the heterogeneous source and destination. The workstations in both of the WiMAX and WiFi network models are configured to facilitate the execution of VoIP applications. The VoIP application, used in this project, is configured to operate as an 'Interactive Voice' service and produce one voice frame per packet. The application profile configuration has been set accordingly in order to make this VoIP application operate in a serial mode. A random generation approach was used to make "Calls" to workstations. The "Calls" were exponentially distributed while having an average duration of three minutes. Furthermore, the call inter-arrival periods are also exponentially distributed. In addition to the application profile and application configuration, the WiMAX network model contains a WiMAX profile. In this profile, a service class of 'Gold' with UGS distribution for VoIP application has been created, which was deployed and classified on all the subscriber stations.

## V.    RESULTS AND DISCUSSION

The average jitter graphs, as shown in Fig. 3(a) and 3(b), were obtained from simulating all three scenarios for one hour. They revealed that WiMAX always has better performance over WiFi.

WiFi also suffered from an extreme level of jitter during the initial five minutes, this was likely because of the nature of the convergence period. Although WiMAX, on the other hand, suffered from a similar hike, it was much lower than that observed for WiFi.

The most interesting result we have found is that the average jitter of WiFi-WiMAX scenario, at some points, exceeds that of WiFi. It should ideally always remain somewhere in-between WiFi and WiMAX. Because the simulation was run based on making random calls and no direct handover was associated, this result is very intriguing. However, further research is required to find out the reason(s) behind such a behavior of the WiFi-WiMAX scenario.

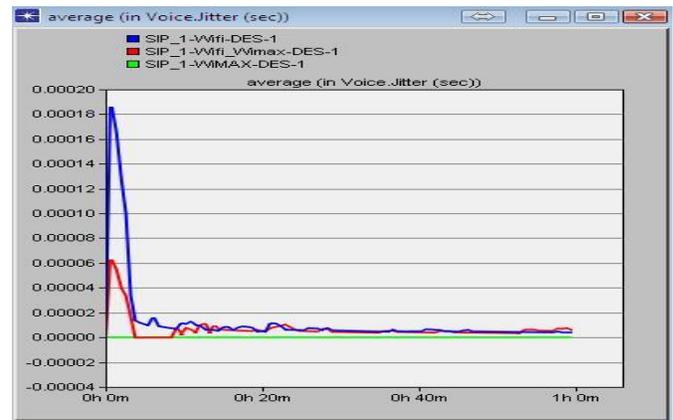

(a)

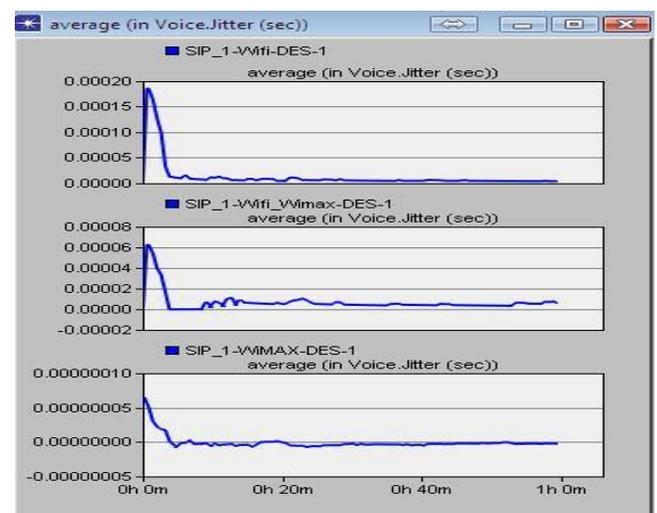

(b)

Fig. 3. (a) Average VoIP Jitter (Overlaid). (Top curve is WiFi, middle curve is WiFi-WiMAX, bottom line is WiMAX). (b) Average VoIP Jitter (Top: WiFi; middle: WiFi-WiMAX; bottom: WiMAX ).

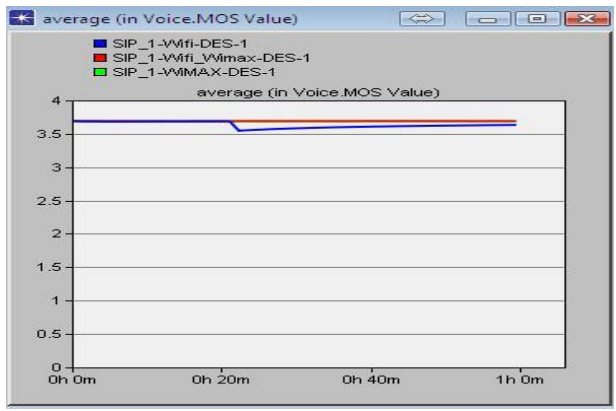

Fig. 4.    Average MOS (Overlaid) of 3.7. (Top line is WiFi, bottom line WiFi-WiMAX).

In terms of the MOS, both WiMAX and WiFi-WiMAX observe similar levels of performance, as shown in Fig. 4. Although the call generation was exponentially distributed, the MOS performance of these two networks remains very steady over the whole simulation period.

On the other hand, although at the beginning of the simulation the WiFi network observes a similar level of MOS. However, as time passes, with the increased level of VoIP traffic due to the higher number of calls generated, the MOS decreases. As a result, taking into consideration the MOS, it can be deduced that both WiMAX and WiFi-WiMAX networks outperform the WiFi network. Moreover, although the MOS of the WiFi-WiMAX network scenario should theoretically be at some mid-point in-between the MOS graphs of WiFi and WiMAX, a much higher performance is observed.

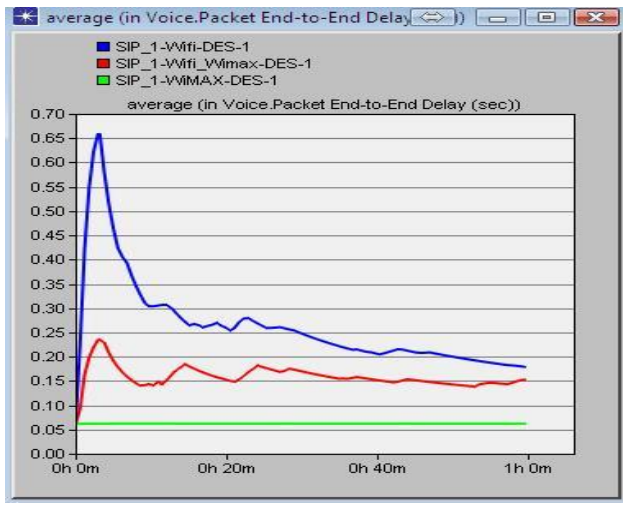

Fig. 5.    Packet End-to-End Delay (Top curve is WiFi, middle curve is WiFi-WiMAX and bottom curve is WiMAX).

With regard to the packet end-to-end delay, WiMAX provides better services in comparison with either using just WiFi or WiFi-WiMAX, as illustrated in Fig. 5. In fact, WiMAX constantly remains in the "Good" range, as outlined in Table 2. Although WiFi observes a high level of packet end-to-end delay at the initial setup phase, it reaches and remains within the "Acceptable" band after the network has converged. The WiFi-WiMAX network remains within the "Acceptable" band even during the convergence period.

## VI.    CONCLUDING DISCUSSION

The paper presented the early findings related to VoIP traffic transmitted through WiFi, WiMAX and WiFi-WiMAX networks. Initially, two scenarios where designed where both generation and termination of the VoIP calls take place in an environment of homogenous networks such as WiFi and WiMAX. Another scenario was later added where calls were generated at the WiFi network and terminated at WiMAX networks and vice-versa.

One of the most thought-provoking findings of our research is regarding the average jitter value of the WiFi-WiMAX scenario of not being in-between WiFi and WiMAX. Our research shows that it does not always perform as expected; even, at some points in time, it exceeds that of WiFi.

The MOS of the WiFi-WiMAX network should ideally be somewhere near halfway of the WiFi and WiMAX MOS graphs. Our research has found that it exhibits a much higher performance than that. Similarly, the packet end-to-end delay of WiFi-WiMAX remains close to that of WiFi and is much higher than expected.

Since there are still a number of WiMAX providers, the study could be strengthened further if comparison between the simulation results against the corresponding results of a real deployment could actually be made. However, due to business and security reasons, companies tend not to reveal their data to the public. If any such data is received, we have plans to compare our results against them.

Future work will include other networks covering: GSM, GPRS, EDGE, UMTS (3G), CDMA, LTE and 4G. The analysis of such QoS parameters for Voice-over LTE (VoLTE) will be one of the particular future research directions. The effect of handover covering, soft, softer and hard on the network traffic will also be focused upon in the future works of this continuing research project.

Furthermore, the scope of this study will be broadened by including the investigation of the impact on other QoS parameters e.g. the packet drop rate, queuing delay and the throughput. To find out the reasons affecting the behavior of these parameters, they will be meticulously examined with the goal of attaining a better optimization and improved efficiency of the network.